\documentclass[lettersize,journal]{IEEEtran}
\usepackage{amsmath,amsfonts}
\usepackage{algorithmic}
\usepackage{algorithm}
\usepackage{array}
\usepackage[caption=false,font=normalsize,labelfont=sf,textfont=sf]{subfig}
\usepackage{textcomp}
\usepackage{stfloats}
\usepackage{url}
\usepackage{verbatim}
\usepackage{graphicx}
\usepackage{cite}
\begin{document}

\title{VeriPy - A New Python-Based Approach for SDR Pipelined/Unrolled Hardware Accelerator Generation}

\author{Yuqin Zhao, Linghui Ye, Haihang Xia,  Luke Seed, Tiantai Deng{

Electronic and Electrical Engineering, the University of Sheffield, Mappin Building, Sheffield, UK, S1 3JD

Corresponding author: Tiantai Deng (email: t.deng@sheffield.ac.uk).

This work is supported by EPSRC Early Career Researcher Studentship
}

\thanks{This paper was produced by the IEEE Publication Technology Group. They are in Piscataway, NJ.}
\thanks{Manuscript received April 19, 2021; revised August 16, 2021.}}

\markboth{Journal of \LaTeX\ Class Files,~Vol.~14, No.~8, August~2021}%
{Shell \MakeLowercase{\textit{et al.}}: A Sample Article Using IEEEtran.cls for IEEE Journals}

\IEEEpubid{0000--0000/00\$00.00~\copyright~2021 IEEE}

\maketitle

\begin{abstract}
Software-defined radio (SDR) plays an important role in the communication field by providing a flexible and custimized communication system for different purposes according to the needs. To enhance the performance of SDR applications, hardware accelerators have been widely deployed in recent years. In facing this obstacle, a necessity arises for a high-level synthesis (HLS) tool specifically designed for communication engineers without detailed hardware knowledge. To lower the barrier between SDR engineers and hardware development, this work proposed a Python-based HLS tool, VeriPy, which can generate both mainstream architecture for hardware accelerators in Verilog specifically for SDR designs including unrolled design and pipelined design, requiring no detailed digital hardware knowledge or Hardware Description Languages (HDL). Furthermore, VeriPy supports automatic testbench generation with random input stimulus, an extensible hardware library, performance and resource estimation, and offers strong optimisation potential at both the algorithmic and digital hardware levels. 

The generated hardware design by VeriPy can achieve up to 70\% faster operating frequency compared to pragma-optimised Vivado HLS designs with a reasonably higher resource consumption while delivering comparable performance and resource consumption to hand-coded implementations. Regarding code complexity, VeriPy requires no pragmas, completely eliminating the need for low-level hardware knowledge. For straightforward algorithms, the input code length remains comparable to that of Vivado HLS.
\end{abstract}


\section{Introduction}
\IEEEPARstart{S}{oftware-defined} radio (SDR) is a communication platform where the signals are processed in the software approach based on a hardware platform with flexible configurations of frequency, bandwidth, etc \cite{bib1}. Conventional SDR applications are operated on a General-Purpose Processor (GPP) like a central processing unit (CPU) \cite{bib1_GPPSDR}, from which the SDR design can benefit from high reprogrammability and short redesign time \cite{bib2}. It demonstrates lower performance-to-power efficiency compared to other platforms, such as Field Programmable Gate Arrays (FPGAs), when applied to an SDR application. This is primarily due to the delay incurred during the execution of software layers and the fetch decoding execution process \cite{bib3}\cite{bib4}. Currently, software-defined radio (SDR) plays an important role in the state-of-the-art 4G and 5G communication systems \cite{bib5_1, bib5_2}. Thanks to its extraordinary flexibility, SDR is projected to have significant potential for development and implementation in future SDR applications experimentally and practically, reflecting the increasing interest in enhancing SDR performance to meet the heavier computational demands of 6G data processing by deploying it on alternative hardware platforms, such as FPGAs, instead of general-purpose processors (GPPs), achieving higher efficiency and lower level of flexibility \cite{bib5_1, bib5_2, bib5, bib6}. For example, future SDR applications in 6G data processing are expected to be computationally intensive on general-purpose processors (GPPs); however, their performance can be significantly enhanced through hardware architectures such as systolic arrays implemented on FPGAs \cite{sysArray}.
		
		To improve the SDR performance-to-power efficiency, FPGA emerges as a more suitable alternative hardware platform to replace the current GPP platform, attributed to its remarkable performance-to-power efficiency \cite{bib6}\cite{bib7}. However, FPGA implementations usually requires extensible hardware knowledge and using the low-level Hardware Description Languages (HDLs), necessitating extensive knowledge and experience in digital hardware, which requires a different skill set from SDR engineers. While general-purpose tools like Vivado HLS have been suggested for SDR hardware development, they still demand considerable hardware expertise \cite{bib8} as HLS only allows coding using a higher level syntax but does not truly bring the design level higher. achieving such optimisations typically requires additional settings and efforts, such as using directives or pragmas alongside the code. Alternatively, users may need to manually modify the HDL code, which often results in poor readability. Both approaches still present a substantial barrier to entry for individuals without extensive hardware knowledge. Alternatively, Vivado can also combine MATLAB Simulink, HDL Coder, lowering the entry barrier for SDR design at the system level design. However, similar to Vivado HLS, it still presents challenges in performing low-level hardware optimisations. Tools such as UHD (USRP Hardware Driver) using GNU Radio for the USRP, which are specifically designed for SDR communication applications, rely on vendors to update corresponding hardware libraries to support software applications and achieve higher-performance functionalities. This dependence imposes a risk for users, as the achievable performance and flexibility may be limited by the availability and updates of these customised libraries \cite{USRP}.
		
		Developing an SDR application using HLS tools like Vivado HLS or Intel Quartus HLS Compiler offers the advantage of directly generating both Verilog and VHDL designs from C/C++ code, supported by comprehensive libraries \cite{IntelHLS, VivadoHLS}. However, this approach also presents several challenges, including the limited generalisability of tool chain (e.g. the output hardware desgin of HLS can be implemented in any FPGA, however, but when it comes to IP in a certain tool chain, the usage of the IP will be limited within the tool chain itself), lack of readability of the generated HDL file, a steep learning curve, and the need for significant hardware expertise. Additionally, hardware-specific pragmas, such as Unrolled/Pipeline and AXI protocol settings (used in HLS C/C++ for coding simplification), introduce barriers for SDR engineers. The choice of different AXI protocols and optimisation pragmas/directives can significantly impact the performance, power, and area of the resulting hardware design. The choice between unrolled and pipelined implementations can lead to different hardware architectures, ultimately affecting performance in terms of timing and resource utilisation. Similarly, selecting different AXI protocols influences the data transfer speed and size in the hardware design. Besides, the optimisation in the vendor's tool, like Intel Quartus HLS and Vivado HLS, relies on the correct placement of directives such as pipeline and unroll pragmas, which increases code complexity and steepens the learning curve \cite{QuatusPragma, VivadoPragma}. Moreover, effective use of these pragmas requires users to possess a solid understanding of hardware design concepts \cite{bib8}. Furthermore, the limited optimisation capabilities are exacerbated by the poor readability of the hardware designs generated by HLS tools, restricting users to optimisation at the HLS pragma/directive level and limiting flexibility for modifications at the HDL level.
		
		GNU Radio, an SDR-specific hardware design tool, accepts Graphical, Python, and C/C++ as input \cite{USRP}. It also benefits from compatibility with various platforms such as GPPs, GPUs, DSP chips, and FPGAs \cite{bib8}. However, the functions supported in GNU Radio are provided by default, with no capability for hardware-level reprogrammability, which limits both the performance and design flexibility of SDR applications \cite{bib9}. Additionally, the transfer speed in a GNU Radio hardware design is limited to USB 2.0 because of the out of date and non-customised library, which falls significantly below the state-of-the-art communication data rates. \cite{bib10}. 
		

		The existing approaches for SDR development on FPGAs typically present one of two limitations: either a steep learning curve and the need for extensive hardware knowledge, or a limited capacity to improve performance for future communication systems. To address these problems and enable SDR engineers to develop FPGA-based applications with a lower hardware entry barrier while achieving an improved trade-off among performance, power, and area (PPA), this paper introduces VeriPy—a novel Python-based, SDR-specific HDL generation tool designed for users without extensible hardware design experience.

        The proposed tool supports the generation of both unrolled and pipelined Verilog hardware to meet different performance, power, and area (PPA) requirements, based on a one-to-one function mapping concept. The Python-based software design is decomposed into smaller functions and transformed into a binary tree that captures data dependencies and timing information. This intermediate representation is then used to construct the corresponding hardware architecture. Additionally, the tool supports automatic testbench generation with random data for input stimulus, an extensible hardware library, performance and resource estimation, and offers strong optimisation potential at both the software and hardware levels.

        To evaluate the effectiveness of VeriPy HLS tool, several benchmarking and commonly-used algorithms from SDR applications \cite{bib12, bib13, bib14} were extracted and implemented using VeriPy, Vivado HLS 2024, and hand-coded Verilog approaches. Based on the results, VeriPy achieves a maximum operating frequency and throughput (they have the same data type) up to 1.7× compared to the pragma-optimised designs generated by Vivado HLS 2024, while delivering performance comparable to hand-coded Verilog implementations. For the trade off, the hardware generated by VeriPy consumes more resources than the HLS version but remains comparable to the hand-coded Verilog implementation.
        
        The main contributions of this SDR hardware generation tool can be summarised below:

		\begin{enumerate}
			\item VeriPy accepts Python as its input design language, which is the most widely used programming language, eliminating the need for extensible hardware knowledge for SDR engineers.

			\item The generated Verilog code includes clear, structured comments for each code block, significantly improving readability and facilitating further hardware-level optimisation.
    
			\item VeriPy supports automatic testbench generation with random input stimulus, an extensible hardware library, and performance and resource estimation—all of which contribute to a faster and more user-friendly design process.
            
            \item The generated hardware achieves up to 1.7× higher operating frequency and throughput in the unrolled FFT design compared to optimised Vivado HLS implementations and performs comparably to hand-coded Verilog, while consuming more resources than the HLS version as trade off but remaining similar to the hand-coded Verilog design.
			
		\end{enumerate}

\section{Tool Structure and Analysis}

    In this section, the VeriPy tool is introduced and analysed from the highest level down to its implementation details. The discussion covers its overall structure and workflow, the data types and fundamental functions it provides, and the detailed internal processing mechanisms. By combining these perspectives, this section aims to provide a comprehensive and clear understanding of VeriPy’s design and functionality.

\subsection{Overview of the VeriPy and Working Flow}

    \begin{figure}
	\centerline{\includegraphics[width=3.5in]{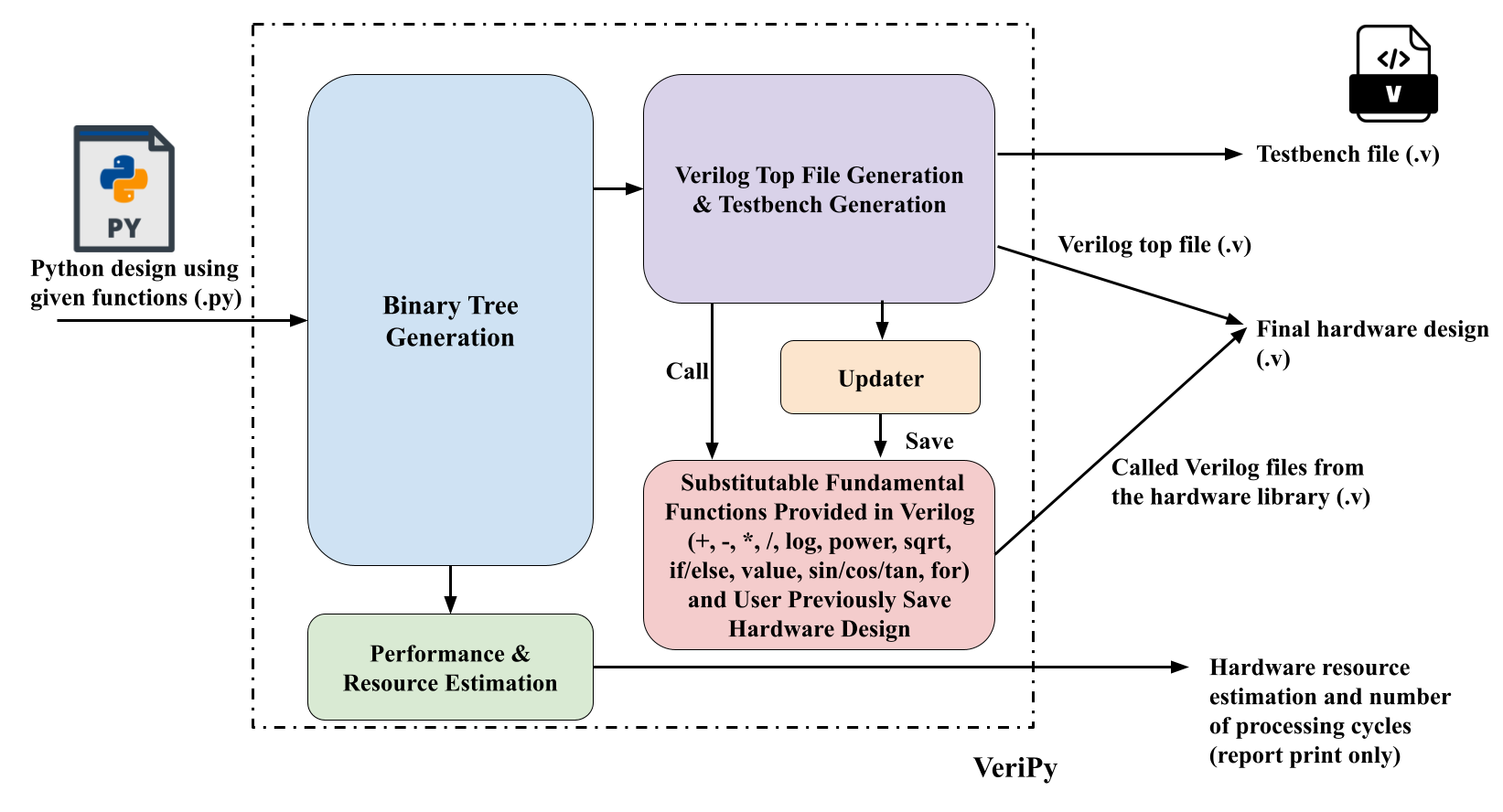}}
	\caption{Tool structure.}
    \label{toolstructure}
    \end{figure}

	The proposed VeriPy is a Python-based high-level synthesis (HLS) tool designed to automatically generate unrolled and pipelined hardware architectures for software-defined radio (SDR) applications without requiring extensive hardware design expertise. The generation environments for both unrolled and pipelined hardware share a common structural framework, as illustrated in Figure \ref{toolstructure}. VeriPy accepts Python-based SDR algorithm code as input, utilising a set of predefined fundamental functions. Based on this input, a binary tree array is constructed during the Binary Tree Generation block, which encapsulates all the essential information needed for hardware synthesis, including operands, operators, delay cycles, and address references. In the Verilog Top File and Testbench Generation block, the necessary information from the binary tree array is extracted to produce a top-level Verilog file that instantiates all required submodules, along with the corresponding testbench file. All data types in the generated files are set as 32-bit fixed-point format, comprising 16 bits for the integer part and 16 bits for the fractional part. The generated top-level Verilog file also includes comments that explain the functionality of each code block, enhancing readability and facilitating hardware-level optimisations. Simultaneously, the Performance \& Resource Estimation block analyses the expected resource usage and processing cycles and generates a report (the performance report is generated for unrolled designs only). Additionally, users can choose to update the hardware library with newly synthesised modules for future reuse.

	 \begin{figure}
		\centerline{\includegraphics[width=3.5in]{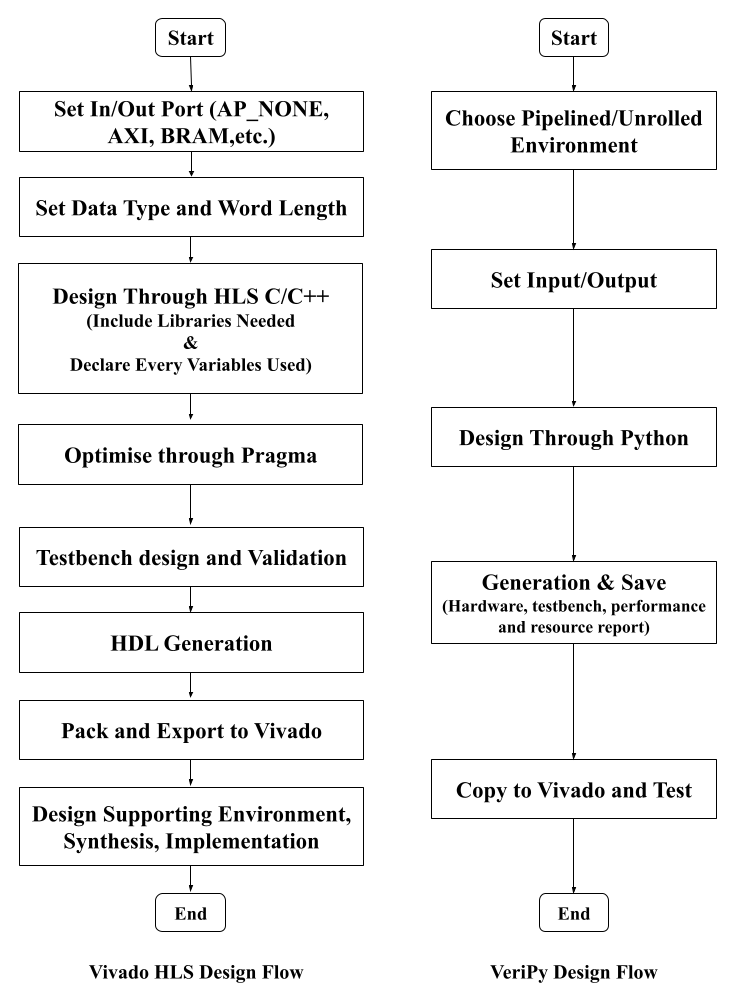}}
		\caption{The  Design Flow Comparison between Vivado HLS and VeriPy.}
        \label{workFlowComp}
	\end{figure}

    A comparison of the design flows of Vivado HLS and VeriPy is presented in Figure \ref{workFlowComp}. Vivado HLS requires a more complex workflow, demanding greater adherence to design rules and a deeper understanding of hardware concepts. In contrast, VeriPy simplifies the process by allowing users to develop hardware designs using only the provided fundamental functions, along with straightforward input and output declarations.
    
    To design an FPGA-based SDR accelerator using Vivado HLS, users must follow several critical steps. First, it is necessary to configure the input/output port types (e.g., AP\_NONE, AXI, BRAM) according to the specific design requirements, which demands a clear understanding of the differences between hardware communication protocols. Careful selection of data types, word lengths, and appropriate libraries is essential, as these choices directly affect the accuracy and performance of the final implementation. Since C/C++ is used as the input language, all variables must be explicitly declared to prevent compilation errors and warnings. In addition, optimisation pragmas must be inserted, which also requires familiarity with hardware design principles. A testbench must be developed to validate the design through C/RTL co-simulation in Vivado, and the verified hardware must then be packaged and exported. Finally, users may need to configure a port-protocol-specific environment within Vivado to complete the synthesis and implementation stages.   
	
    VeriPy, on the other hand, offers a significantly streamlined design flow, particularly when compared to Vivado HLS. Users begin by selecting either the pipelined or unrolled environment, depending on their specific requirements. During the Python algorithm development stage, users will define inputs and outputs and construct the desired algorithm exclusively with the provided function set. Once the design is complete, VeriPy automatically generates the corresponding hardware description and its testbench in Verilog, with all data defined using a 32-bit fixed-point data type (16 bits for the integer part and 16 bits for the fractional part), along with a detailed performance and resource utilisation report. Additionally, the generated hardware modules are automatically stored in the hardware library for future reuse. Finally, users can import both the hardware design and testbench into Vivado to perform functional validation, synthesis, and implementation.

    \begin{figure}
		\centerline{\includegraphics[width=3.5in]{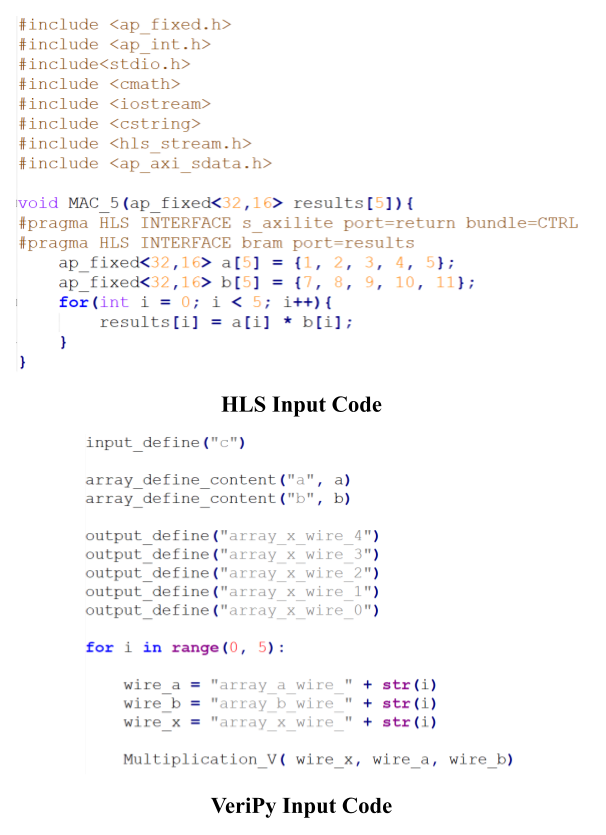}}
		\caption{Comparison of the Input Code between HLS and VeriPy.}
        \label{InputCodeComp}
	\end{figure}

    Figure \ref{InputCodeComp} illustrates a comparison of manually written input code for both approaches. VeriPy significantly reduces the complexity associated with variable declarations and library inclusions required by Vivado HLS, thereby eliminating potential confusion for SDR engineers and streamlining the design process.

   Compared with Vivado HLS, the design stages in VeriPy are significantly simplified and streamlined. Variable declarations and explicit data type specifications are not required, and hardware-specific pragmas for optimisation are unnecessary, as VeriPy directly provides dedicated Python-based unrolled and pipelined generation environments. Additionally, a corresponding testbench is automatically generated, further reducing the time required for testbench development. In summary, VeriPy offers clear advantages in terms of shorter design time, ease of use, minimal hardware expertise requirements, and reduced coding effort.

\subsection{Datatypes and Fundamental Functions}
    
    As mentioned previously, to lower the barrier for users with limited hardware expertise, VeriPy does not require explicit data type declarations, as all data is by default defined as 32-bit fixed-point numbers, with 16 bits for the integer part and 16 bits for the fractional part in the generated hardware. Since SDR systems typically receive input from an ADC with a resolution of 4 to 14 bits \cite{bib15}, the 32-bit fixed-point representation adopted in this work is expected to be sufficient. This is implemented through the provided fundamental functions, as all corresponding hardware modules for these functions are designed using the 32-bit fixed-point format.

    \begin{table}
    \centering
    \caption{Fundamental Functions and Their Expression}
    \begin{tabular}{|>{\centering\arraybackslash}p{4cm}|
                    >{\centering\arraybackslash}p{4cm}|}
			\hline
			{Functions}&{Expression}\\
			\hline
			
			Addition\_V(c,a,b)&c = a + b\\
            \hline
			Subtraction\_V(c,a,b)&c = a - b\\
            \hline
			Multiplication\_V(c,a,b)&c = a * b\\
            \hline
			Division\_V(c,a,b)&c =  a / b\\
            \hline
			Power\_V(c,a,d)& $c = a^d$\\
            \hline
			Logarithm\_V(c,a,b)&c = log(a)b\\
            \hline
			Sqrt\_V(c, a)&c = sqrt(a)\\
            \hline
			SinCosTan\_V(sin\_r, cos\_r, tan\_r, a)&sin\_r = sin(a), cos\_r = cos(a), tan\_r = tan(a)\\
            \hline
			Value\_V(c, a)&c = a\\
			\hline
            If\_V(comp1, comp2, condition) *& if(comp1 (condition) comp2)\\
            \hline
        If\_V(comp1, comp2, condition, input\_array, output\_array) **&if( comp1 (condition) comp2 )\\
    \hline
    \multicolumn{2}{p{8cm}}{* If/else in unrolled environment ** If/else in pipelined environment. input\_array and output\_array are for input/output declaration in the if/else block, they are not used in the calculation.}

    \end{tabular}
    
    \label{FundFunExpression}
\end{table}

\begin{table}
    \centering
    \caption{Input and Output Precision of Provided Fundamental Functions.}
    \begin{tabular}{|>{\centering\arraybackslash}p{3cm}|
                    >{\centering\arraybackslash}p{2cm}|>{\centering\arraybackslash}p{2cm}|
              }
    \hline
       {Functions}&{Input Precision}&{Output Precision}\\
       \hline
			
	Addition\_V(c,a,b)&fixed(32,16)&fixed(32,16)\\
    \hline
	Subtraction\_V(c,a,b)&fixed(32,16)&fixed(32,16)\\
    \hline
	Multiplication\_V(c,a,b)&fixed(32,16)&fixed(32,16)\\
    \hline
	Division\_V(c,a,b)&fixed(32,16)&fixed(32,16)\\
    \hline
	Power\_V(c,a,d)&fixed(32,16) int(8)*&fixed(32,16)\\
    \hline
	Logarithm\_V(c,a,b)&fixed(32,16)&fixed(32,16)\\
    \hline
	Sqrt\_V(c, a)&fixed(32,16)&fixed(32,16)\\
    \hline
	SinCosTan\_V(sin\_r, cos\_r, tan\_r, a)&fixed(32,16)&fixed(32,16)\\
    \hline
	Value\_V(c, a)&fixed(32,16)&fixed(32,16)\\
    \hline
	If\_V(comp1, comp2, condition)**&fixed(32,16)&bool(1)***\\
    \hline
        If\_V(comp1, comp2, condition, input\_array, output\_array)****&fixed(32,16)&bool(1)\\
	\hline
        \multicolumn{3}{p{8cm}}{* Only support power(a,b) when b is an integer from -128 to 127. ** Unrolled if/else. *** If/else only gives true/false as output. **** Pipelined if/else. }

    \end{tabular}
    
    \label{tab:Ch4FundamentalFunPrecision}
\end{table}

\begin{table}
    \centering
    \caption{Fundamental Functions Internal Precision}
    \begin{tabular}{|c|c|}
			\hline
			{Functions}&{Internal Precision}\\
			\hline
			
			Addition\_V(c,a,b)&fixed(32,16)\\
            \hline
			Subtraction\_V(c,a,b)&fixed(32,16)\\
            \hline
			Multiplication\_V(c,a,b)&fixed(64,32)\\
            \hline
			Division\_V(c,a,b)&fixed(64,32)\\
            \hline
			Power\_V(c,a,d)&fixed(64,32)\\
            \hline
			Logarithm\_V(c,a,b)&fixed(128,64)\\
            \hline
			Sqrt\_V(c, a)&fixed(16,8)\\
            \hline
			SinCosTan\_V(sin\_r, cos\_r, tan\_r, a)&fixed(32,29)\\
            \hline
			Value\_V(c, a)&fixed(32,16)\\
			\hline
    \end{tabular}
    
    \label{tab:Ch4FundFunInnerPre}
\end{table}

    Table \ref{FundFunExpression}, Table \ref{tab:Ch4FundamentalFunPrecision}, and Table \ref{tab:Ch4FundFunInnerPre} present the provided fundamental functions, their input/output precision, and their internal calculation precision, respectively. The available fundamental functions include addition, subtraction, multiplication, division, exponentiation, logarithm, square root, sine, cosine, tangent, value assignment, and conditional branching (if/else). Loop structures can be implemented directly using Python’s native for loop syntax. All provided functions support 32-bit fixed-point input and output precision, except for the power function and the if/else block. Currently, the power function supports only an exponent specified as a signed 8-bit integer, since fractional exponents can cause excessive power and resource consumption, and large exponents can easily lead to overflow. For the if/else block, the output is limited to a boolean value, as it performs comparison operations only. Regarding internal precision, addition, subtraction, and value assignment use fixed(32,16), while power, multiplication, division, logarithm, and trigonometric functions (sine, cosine, tangent) use a higher-precision data type to improve numerical accuracy. The square root function, on the other hand, employs an internal precision of fixed(16,8) due to its specific mathematical calculation method. The set of fundamental functions is sufficient for typical SDR algorithm design, and the hardware library will continue to expand as users and developers contribute additional modules over time. This ensures that future designs become increasingly straightforward and convenient, with a wider range of functions available, thanks to VeriPy’s built-in updater and extensible hardware library.

	 \begin{figure}
		\centerline{\includegraphics[width=3.5in]{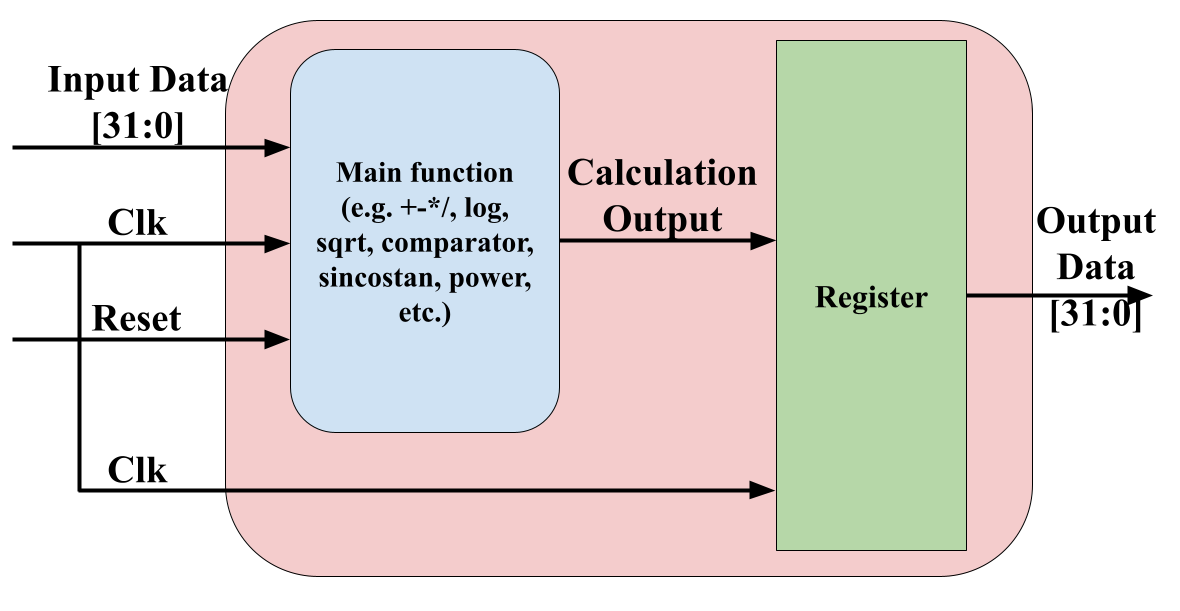}}
		\caption{Hardware Architecture of the Provided Fundamental Functions.}
        \label{FundamentalArchi}
	\end{figure}

\begin{figure}
		\centerline{\includegraphics[width=3.5in]{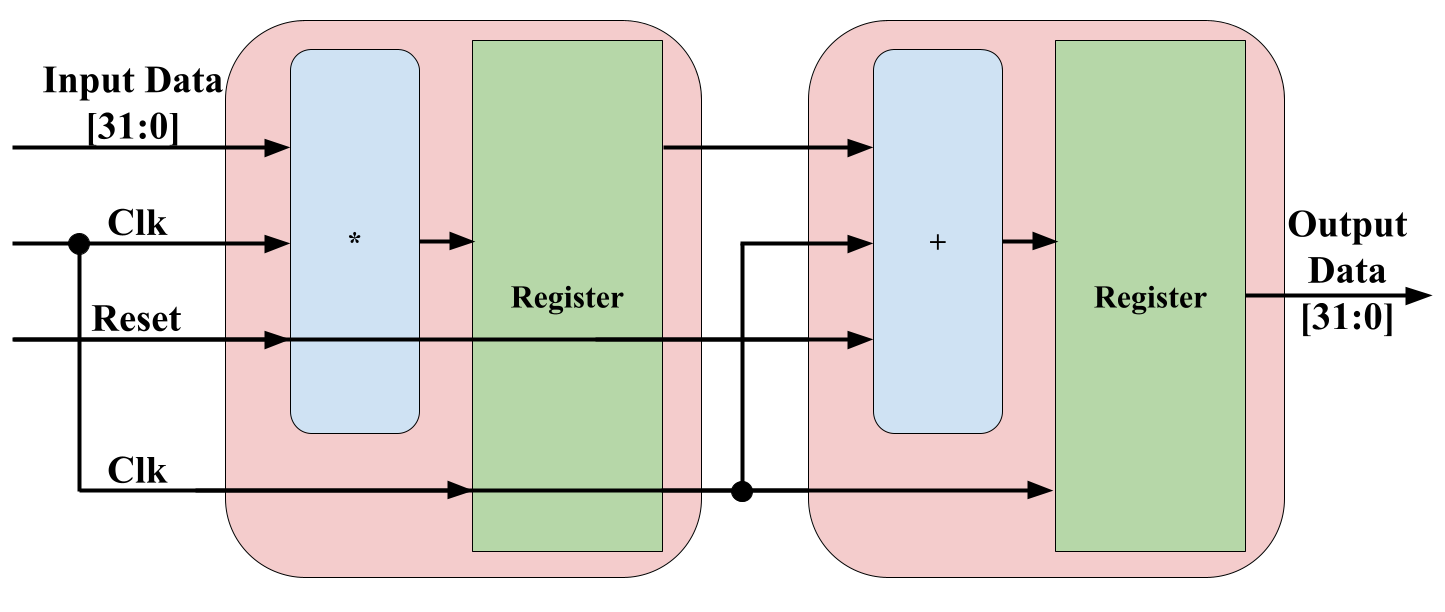}}
		\caption{Generated Unrolled Hardware Architecture Composed of Provided Fundamental Functions.}
        \label{Ch4ArichitectureUnrolled}
	\end{figure}

\begin{figure}
		\centerline{\includegraphics[width=3.5in]{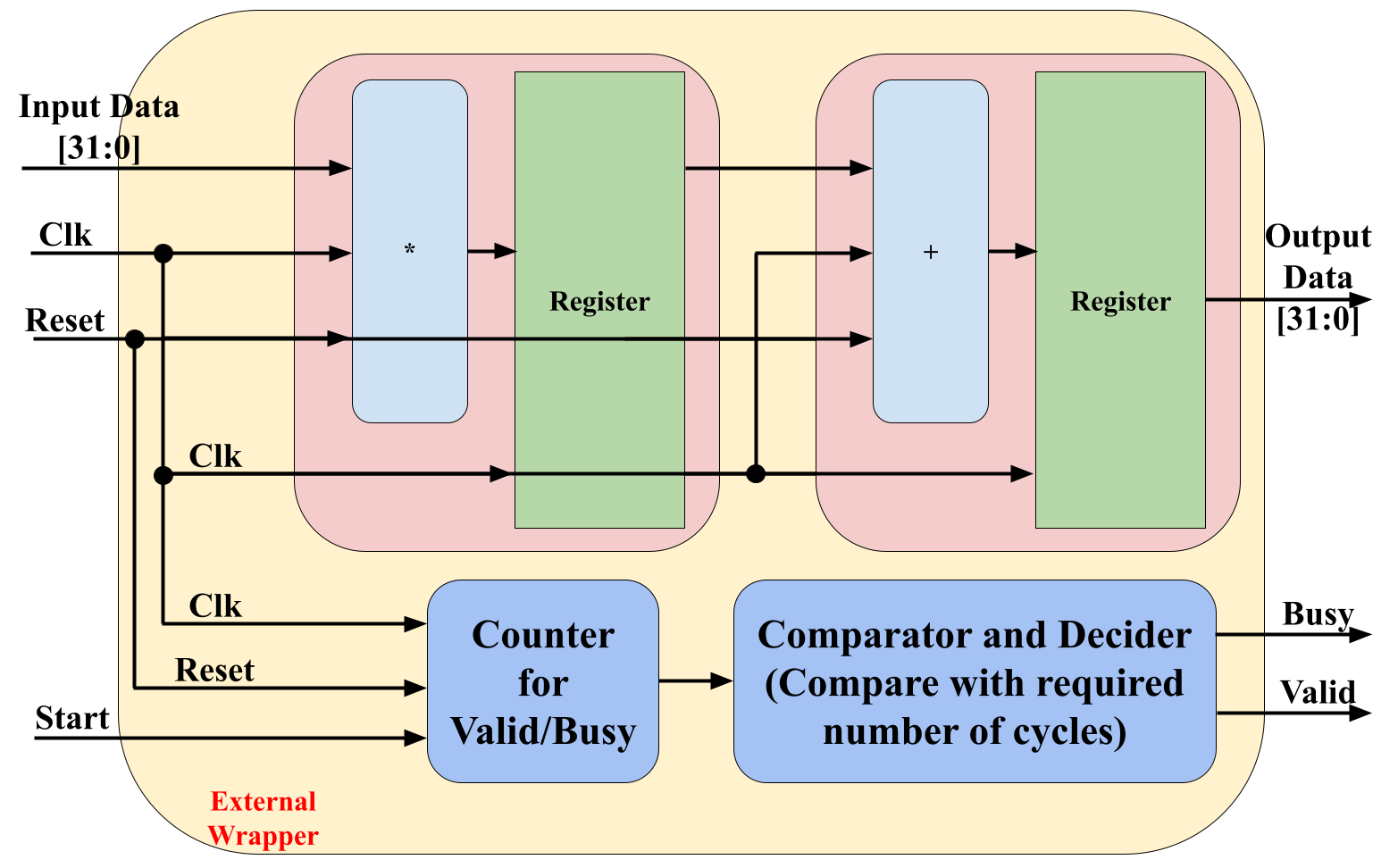}}
		\caption{Generated Pipelined Hardware Architecture Composed of Provided Fundamental Functions.}
        \label{Ch4ArichitecturePipe}
	\end{figure}

     All provided functions share a similar hardware architecture, with signals including input, clock (clk), and reset, and each ending with a register to store the result for the next computation stage, as illustrated in Figure \ref{FundamentalArchi}. Within this architecture, all fundamental functions are synchronised by the clock signal and support reset functionality. Since both the unrolled and pipelined hardware designs generated by VeriPy are composed of these fundamental functions, the resulting architecture is naturally divided into stages, with registers placed between each function to store intermediate results, as shown in Figure \ref{Ch4ArichitectureUnrolled}. The pipelined design, in contrast, is wrapped with control signals such as start, busy, and valid, enabling control flow management through a counter and comparator based on the defined number of processing cycles, as illustrated in Figure \ref{Ch4ArichitecturePipe}.

    It is important to note that the default fundamental functions provided by VeriPy can be directly substituted to meet different design considerations, such as performance or power optimisation. This flexibility enables a potential optimisation approach within VeriPy, as the generated top-level file serves as a framework that instantiates all required submodules. To replace a fundamental function, users simply need to maintain the same module name and port definitions in their custom implementation. Further details on this substitution mechanism will be presented in the next section.

\subsection{Updater and Hardware Library}

\begin{figure}
		\centerline{\includegraphics[width=3.5in]{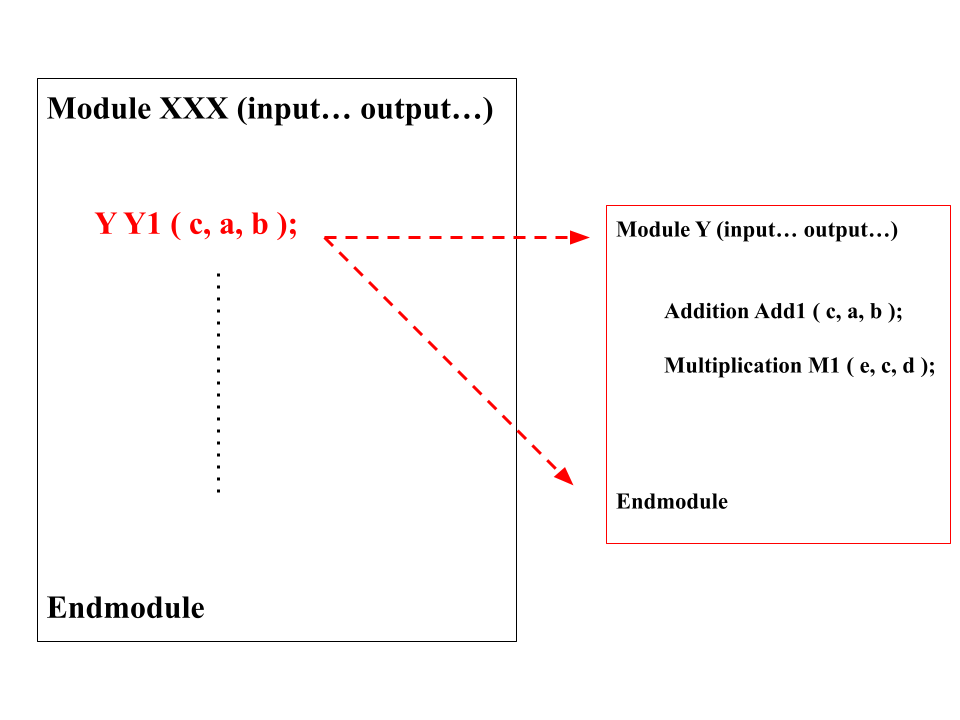}}
		\caption{Nested Structure of the Output Hardware Design.}
        \label{NestedStructure}
	\end{figure}

	The updater function saves the number and order of inputs and outputs for each hardware design, along with the required processing cycles and resource usage, into the hardware library, assigning a unique label to each design. Simultaneously, three Python files corresponding to the If, Else, and Normal statements are automatically generated. When the function is invoked in the user’s Python design, the algorithm first searches the hardware library using the label, ensuring that variable names follow the same order as the declared inputs and outputs. This information is then stored in a binary tree, which is subsequently used during the generation of the Verilog hardware design. The generation process follows the detailed procedure described in the next section. The output maintains a nested structure, where the top-level Verilog file generated by VeriPy directly calls the submodules containing the corresponding hardware designs, as illustrated in Figure \ref{NestedStructure}.

    Through the Updater and Hardware Library, users can design and reuse hardware modules without the need for redesign. This not only saves redevelopment time but also gradually enriches the library, making it more comprehensive and versatile.

\section{Process of Generation}

\begin{figure}
		\centerline{\includegraphics[width=3.5in]{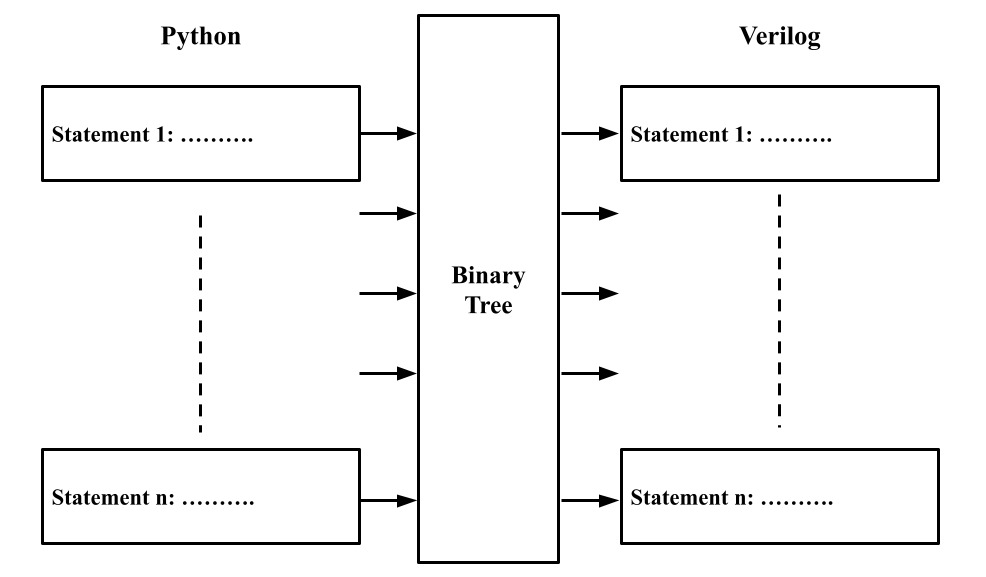}}
		\caption{One-to-one correspondence concept.}
        \label{onetoone}
	\end{figure}

    As illustrated in Figure \ref{onetoone}, the core principle of VeriPy is built upon a one-to-one correspondence between software functions and their hardware counterparts. The functions available in Python are exactly matched to equivalent modules in the hardware library, ensuring a direct mapping. The Python input for VeriPy is composed using these provided functions, each of which is extracted along with its data dependencies and stored in a binary tree structure. Subsequently, the generated Verilog hardware invokes the same corresponding modules from the hardware library and establishes the appropriate wire connections to ensure correct functionality.

    The entire code generation process in VeriPy consists of two main stages: Binary Tree Generation and Hardware Generation. While both the unrolled and pipelined environments share the same Hardware Generation stage, they differ in the Binary Tree Generation stage. This is because pipelined hardware requires more intricate dataflow control and precise ordering of operations, making its binary tree structure more complex than that of the unrolled design. To accommodate these additional requirements, the Binary Tree Generation process is further divided into two phases: Initial Binary Tree Generation, which produces a tree structure sufficient for unrolled hardware, and Binary Tree Processing, which augments the tree with the additional information necessary for pipelined hardware generation.

    \subsection{Binary Tree Generation}
\begin{figure}
        \centerline{\includegraphics[width=3.5in]{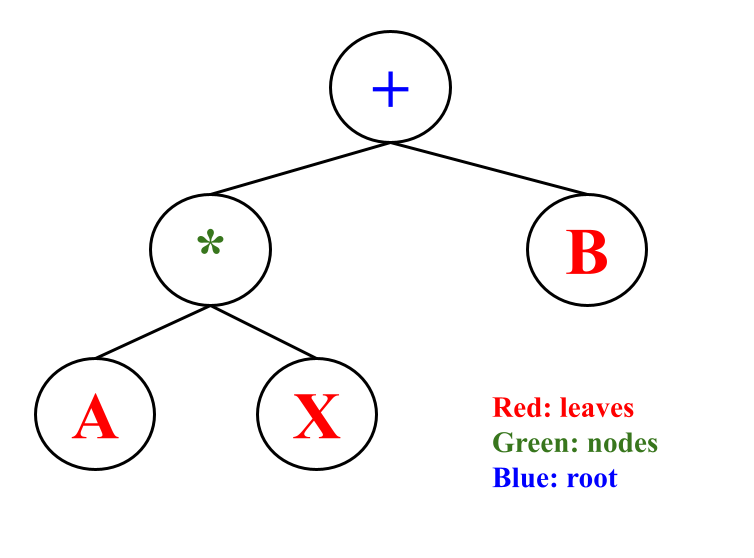}}
	\caption{Binary tree for a simple equation.}
    \label{BinaryTreeEx}
    \end{figure}

    The binary tree concept is employed in VeriPy to provide a clear representation of calculation order and data dependencies. As illustrated in Figure \ref{BinaryTreeEx}, the binary tree for a simple equation Y=A*X+B demonstrates the roles of leaves, nodes, and the root: leaves represent operands, nodes denote operators, and the root corresponds to the final operation of the expression. The original equation can be reconstructed by traversing the binary tree from the leaves to the root. In practice, the binary tree serves as an intermediate representation (IR) between the input and output languages. The design specified in the input language is first transformed into a binary tree, which is then traversed and interpreted in reverse to generate the corresponding hardware design in the output language.

    \subsubsection{Initial Binary Tree Generation}

    \begin{enumerate}
        \item \textbf{Binary Tree Structure in Python}\\ 
		Owing to its structure and functionality, the binary tree also serves as a crucial intermediate representation (IR) of the input Python-based SDR application in VeriPy. The binary tree is internally constructed using an object-oriented programming approach (not exposed to the user). This internal structure consists of eight object fields that store the following information: operand names, number of operands, operator type, result names, number of results, previous addresses, number of previous addresses, and delay cycles.\\

		\item \textbf{Binary Tree Generation Example} \\
		One of the examples is shown in Figure \ref{fig7}, which contains some fundamental functions and if/else. Information at addresses 2, 4, and 5 in the binary tree will be explained. \\
		For address 2, according to the equation A=C*D, [`C', `D'] indicates the operands, 2 indicates there are 2 operands, `A', and `1' mean the only result name `A'. Besides, [-1,1] specifies there is no previous address for `C', and `D' has a previous address at address 1 while [1,1] represents both `C' and `D' have only 1 previous address. The final 1 indicates this multiplication will have a 1-cycle delay for calculation. \\
		Turning to address 4, the equation of A=0 is added to the else part to simplify the search previous address process. It also eliminates the complexity of the binary tree.\\
		Address 5, however, stores a [(4,4),-1] in the object of ``previous addresses'' as A shows up both in the if and else statement, which also gives a [2,1] in the object of ``the number of previous addresses''.
		\begin{figure}
			\centerline{\includegraphics[width=3.5in]{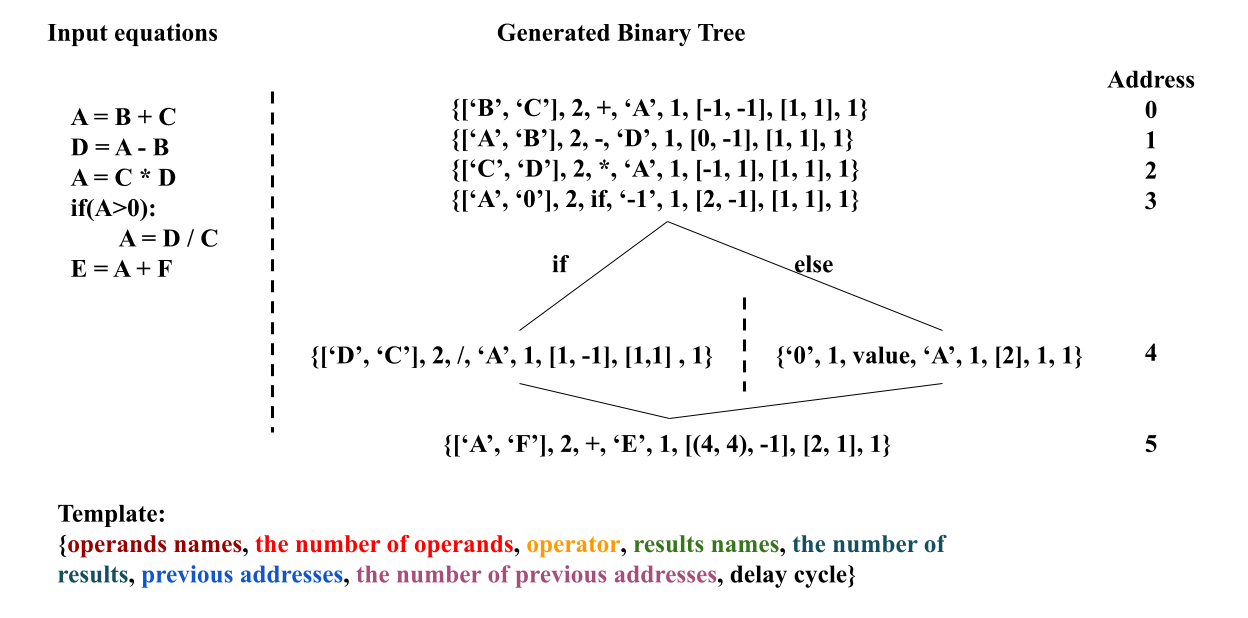}}
			\caption{Binary tree structure and a binary tree generation example.}
            \label{fig7}
		\end{figure}

	\end{enumerate}

    \subsubsection{Processed Binary Tree Generation}

    \begin{figure}		
        \centerline{\includegraphics[width=3.5in]{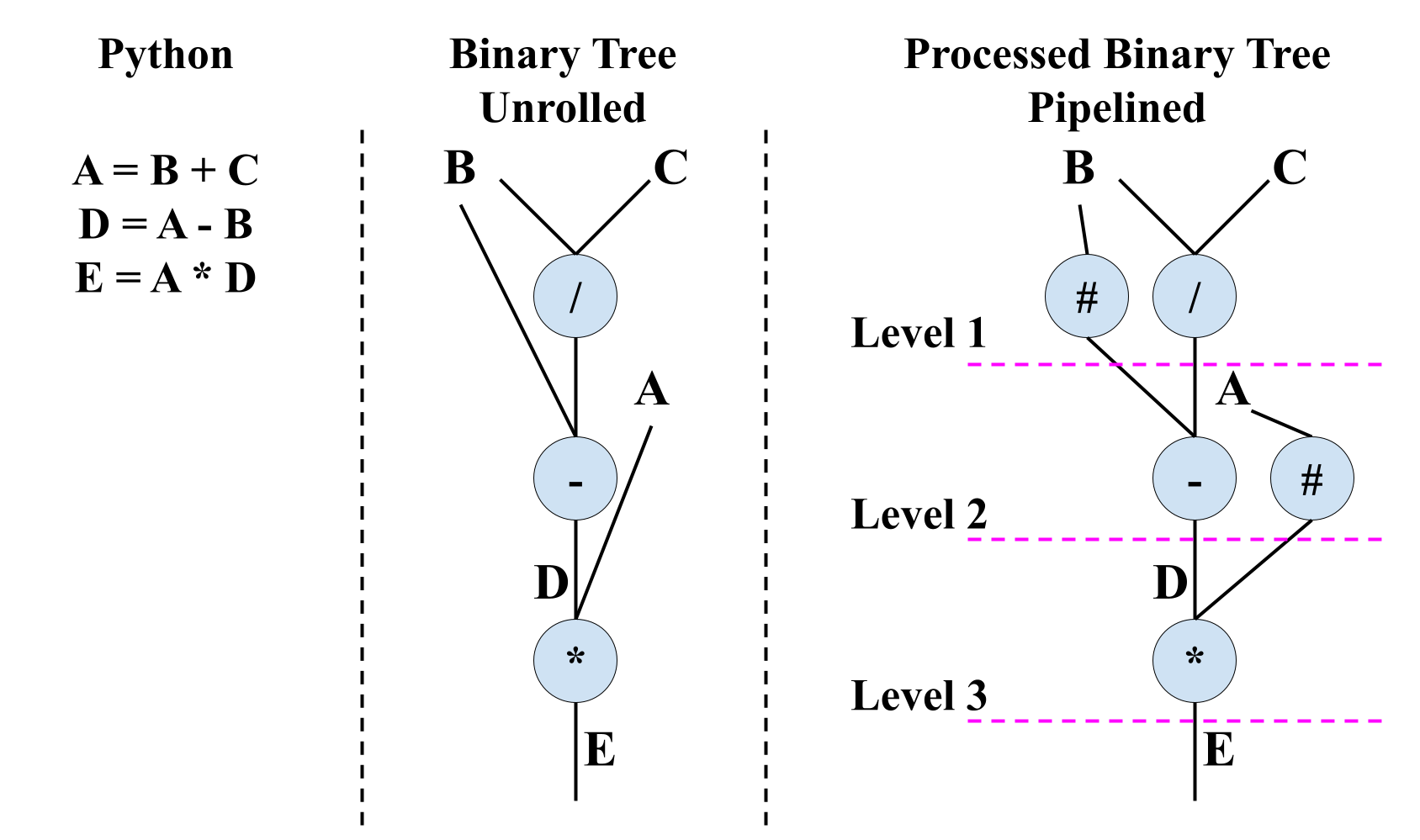}}
	\caption{Unrolled and Pipelined Binary Tree (\# represents delay).}
    \label{PipelineandUnroll}
    \end{figure}

    As shown in Figure \ref{PipelineandUnroll}, the initially generated binary tree in VeriPy contains only data dependency and operator/operand/result information, which is sufficient for unrolled hardware generation. However, additional processing is required to adapt the binary tree for pipelined hardware generation, as pipelining demands precise control of dataflow and execution order across processing stages. This adaptation involves inserting delay modules between stages to align data correctly and ensure proper timing for data propagation.

    Based on the initial binary tree, the complete processing sequence can be extracted and organised into hierarchical levels, representing the operations to be performed at each pipeline stage. To enable pipelined hardware generation, delay modules are inserted between result outputs and their corresponding input ports whenever a level gap exists between dependent operations.
    
    As illustrated in Figure \ref{PipelineandUnroll}, the initial binary tree (centre) displays data dependencies, connections, and required modules—information sufficient only for unrolled hardware generation. To adapt this for pipelined hardware, the dataflow must be explicitly structured and stage boundaries clearly defined. In the example, operations are divided into three levels based on their calculation order. For instance, variable B is required in both Level 1 and Level 2, while variable A, computed as the result of B + C at the end of Level 1, is also needed as input to a multiplication operation in Level 3. Consequently, delay modules are inserted: one between B and the subtraction input in Level 2, and another between the result A from Level 1 and the multiplication input in Level 3. These inserted delays ensure correct data alignment and synchronisation across pipeline stages.

    \subsection{Verilog Hardware Generation}

    The unrolled or pipelined Verilog hardware design is generated by VeriPy based on the initial or processed binary tree, respectively. The binary tree structure stores essential information such as input names, output names, number of inputs and outputs, previous addresses, number of previous addresses, and the number of processing cycles. For the processed binary tree used in pipelined hardware generation, additional delay modules are inserted where necessary; however, no new types of information are introduced—only timing alignment is modified.

\begin{figure}
	\centerline{\includegraphics[width=3.5in]{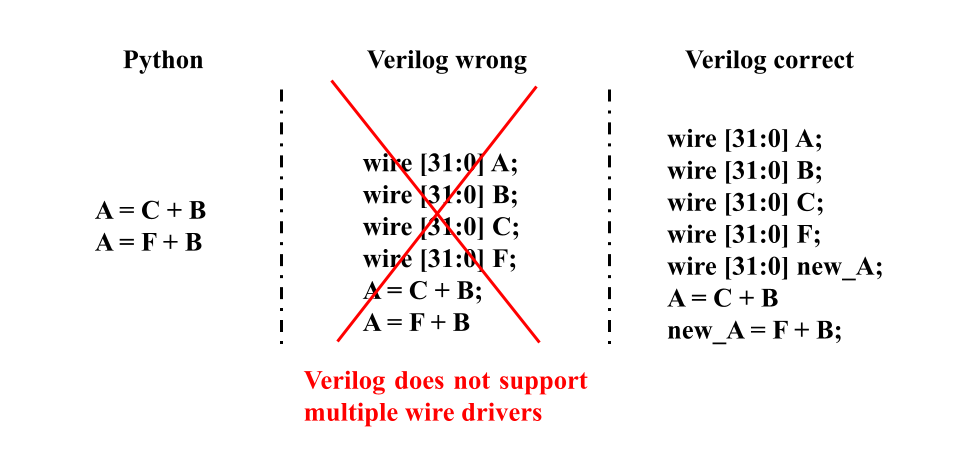}}
	\caption{Different rules in Python and Verilog.}
        \label{HWSW_NewVariable}
    \end{figure}

    As is well known, hardware description languages (HDLs) like Verilog differ fundamentally from high-level programming languages such as Python. Verilog explicitly describes hardware structures, including the creation of wires and registers for data transmission and storage, whereas in Python, variable declarations merely allocate memory without hardware semantics. For example, as shown in Figure \ref{HWSW_NewVariable}, a variable reassignment in Python is interpreted as a new memory allocation, but in Verilog, it could result in multiple drivers assigned to the same wire, leading to synthesis warnings or functional errors. To address this, VeriPy ensures that a new wire is defined for each reassigned signal, thereby avoiding multi-driver conflicts in the generated hardware.

\begin{figure}
		\centerline{\includegraphics[width=3.5in]{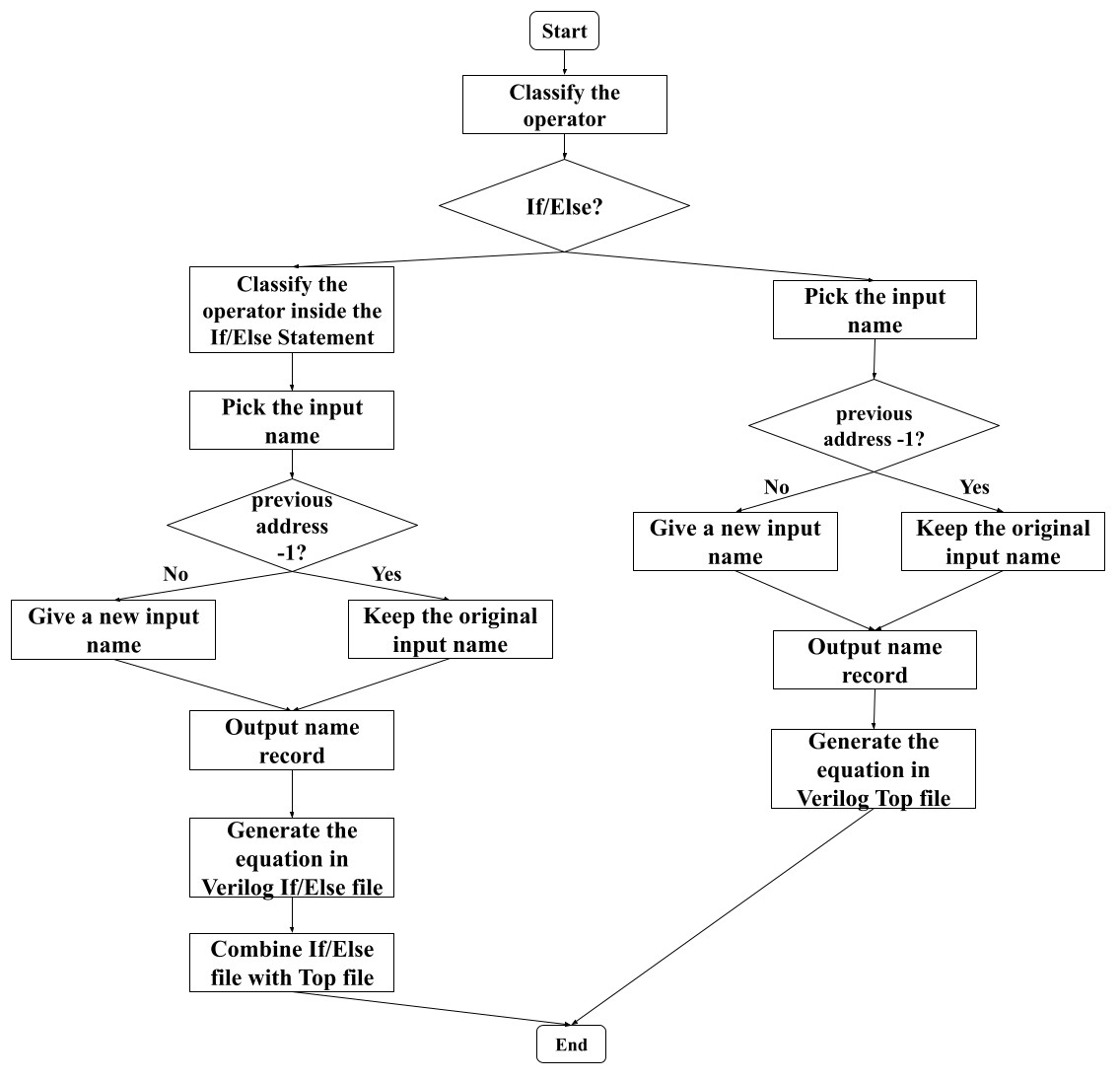}}
		\caption{Flow chart of Verilog generation through binary tree.}
        \label{VerlogGenFlow}
	\end{figure}

	To generate a synthesisable Verilog hardware design, the process of binary tree analysis and code generation must be executed with care. The workflow for Verilog generation is illustrated in Figure \ref{VerlogGenFlow}. 
	
	This process will occur sequentially based on the binary tree generated from the Python design. Since If/else statements will increase the number of branches of the binary tree, the equation in the If/else statement will be processed separately from the ordinary equations. Ordinary equations will be generated directly into the Verilog Top file while equations in the If/else statement will be generated into an If/Else Verilog file. 
	When processing each element, the operator will initially be classified to determine whether it represents an If/Else or a non-If/Else statement. If the current element is a non-If/Else statement, the input name will be selected, and its iteration in the previous occurrences will be checked. A new input name will be assigned if the same name has been used before, while the name will remain unchanged if it hasn't. This process helps avoid errors and warnings in the Verilog design, as mentioned in Figure \ref{HWSW_NewVariable}. The output name will also be recorded for future processes, and the equation with new/original input names (depending on iteration) will be written in the Verilog Top file. 
	If the current element is an If/Else statement, the process will traverse the If/Else binary tree, following the same operations as in the non-If/Else scenario. However, the final equation will be generated in the If/Else file in Verilog. Ultimately, the If/Else file will be combined with the Verilog file to complete the synthesis process.

\begin{figure}
		\centerline{\includegraphics[width=3.5in]{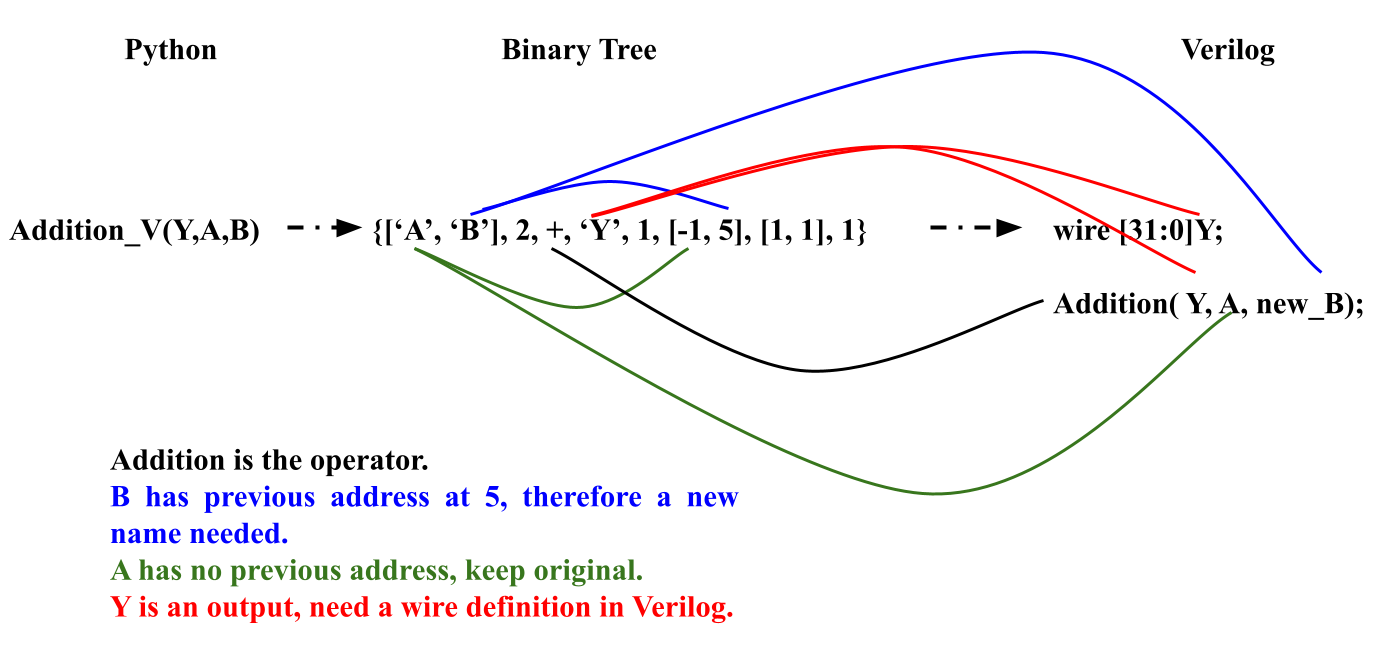}}
		\caption{Example of the process from Python to Verilog design.}
        \label{fig9}
	\end{figure}

	An example is illustrated in Figure \ref{fig9}. A newly declared variable Y is computed as the sum of A and B. Upon transformation into a binary tree, it is revealed that A does not have a previous address, whereas B has a previous address at 5. Consequently, A can retain its original name, while B must be assigned a new name to prevent the occurrence of the multi-driver problem in Verilog. After declaring the new output Y and rewriting the Verilog equation with A and new\_B, the final Verilog design is shown on the right side of Figure \ref{fig9}.

\section{Functions and Rules of VeriPy}

    \subsection{Functions and Advantages of VeriPy}
	
    The key functions and advantages of VeriPy over Vivado HLS are summarised below:

    \begin{enumerate}
        
        \item \textbf{Python Input Without Pragmas or Data Type Declarations}\\
        As illustrated in Figure \ref{InputCodeComp}, VeriPy eliminates the need to declare variables, specify data types, define ports, or insert pragmas for optimisation—unlike Vivado HLS. This significantly reduces the learning curve and lowers the barrier between SDR software engineers and hardware design.

        \item \textbf{Unrolled and Pipelined Hardware Generation}\\
        VeriPy provides two independent environments for unrolled and pipelined hardware generation, catering to different performance, power, and area (PPA) requirements. Unlike Vivado HLS, which relies on pragma directives, VeriPy separates the two modes explicitly to minimise design confusion.

        \item \textbf{Enhanced Readability}\\
        In the generated hardware file, each block of Verilog code is followed by comments explaining its functionality, significantly improving readability—especially in comparison with the Vivado HLS-generated Verilog, as shown in Figures \ref{GenCodeVeriPy} and \ref{GenCodeHLS}.

        \begin{figure}
		\centerline{\includegraphics[width=3.5in]{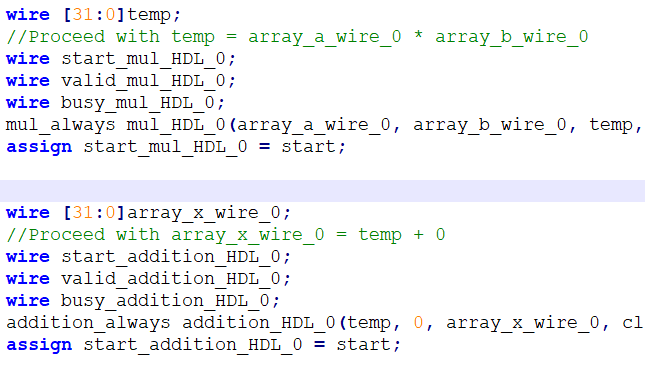}}
		\caption{VeriPy Generated Verilog File.}
        \label{GenCodeVeriPy}
	\end{figure}

        \begin{figure}
		\centerline{\includegraphics[width=3.5in]{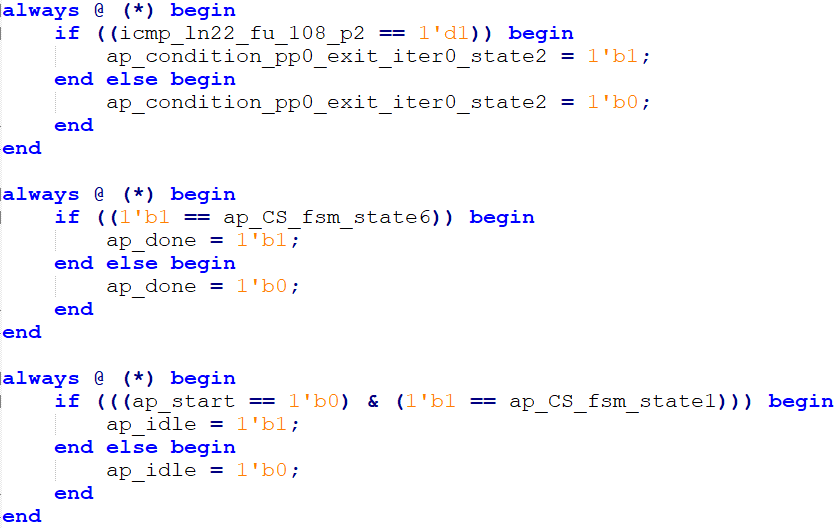}}
		\caption{Vivado HLS Generated Verilog File.}
        \label{GenCodeHLS}
	\end{figure}

       \item \textbf{Support for Both High-Level and Low-Level Optimization}\\
        VeriPy supports optimisation at both the high-level (Python code) and low-level (Verilog structure). The advanced readability of the generated hardware—enhanced by informative comments and a clear structural hierarchy in the top-level Verilog file—enables users to substitute individual fundamental functions or adjust the hardware structure manually with ease.

        \item \textbf{Automatic Testbench Generation}\\
        VeriPy automatically generates a corresponding testbench for each hardware design. This reduces the manual effort required for testbench development and allows users to directly import both the design and testbench into hardware design tools for simulation and validation.

        \item \textbf{Performance and Resource Estimation}\\
        VeriPy provides resource utilisation estimates—including LUTs, flip-flops (FFs), DSPs, and BRAMs—for both unrolled and pipelined designs. Additionally, it offers performance estimation (in terms of required processing cycles) for pipelined designs.

	\item \textbf{Extensible Hardware Library}\\
        VeriPy supports an extensible hardware library where users can store custom hardware modules for future reuse. This promotes design reuse and reduces both development time and input code length.

	\end{enumerate}

	\subsection{Rules}
	
	To generate synthesizable Verilog output, the VeriPy tool requires users to follow several specific rules, as outlined below (for details rules: https://github.com/RainChinChao/PythonToVerilogTool):

		\begin{enumerate}
		\item \textbf{In/Out Define}\\ 
		To ensure the hardware design is synthesizable, input and output names must be defined using the  input\_define() and output\_define()  functions. Currently, the number of input and output names is limited to a maximum of 20 each.

		\item \textbf{Number to Fixed Type} \\
		Since this tool supports a default 32-bit data type (16-bit integer, 16-bit fraction) for inputs and outputs, any pure number must be converted to a 32-bit fixed-point format using number\_to\_hex().

		\item \textbf{Function Output Names and Accumulation} \\
		In this tool, input and output names must be unique within a single function. For accumulation designs, use a new name for each accumulation step. (Examples are available in the instructions on GitHub.)
		
		\item \textbf{Nested If/Else} \\
		Currently, nested if/else structures can be implemented up to 2 levels. For designs requiring additional levels of nesting, please use the updater instead.
		
		\item \textbf{Array} \\
		In this tool, arrays can be declared but will follow a specific naming convention. For instance, if an array A is declared, each element in A will be named array\_A\_wire\_i, where i represents the index of the element.
		
	\end{enumerate}

	Detailed rules, examples, function instructions, and example videos can be found on the following GitHub Repo (https://github.com/RainChinChao/PythonToVerilogTool).

\section{Case Study and Evaluation}
		
		In this section, 5 major SDR examples are chosen for hardware developments of VeriPy, Vivado HLS 2024.1, and handcoded Verilog based on the FPGA platform of RFSoC. Both tools have an intentionally unoptimised SDR algorithm as input to emulate the unfamiliarity between SDR engineers and hardware development. The comparisons are made under the scopes of \textbf{performance \& area}, \textbf{readability \& substitutability}, and \textbf{length of code (LoC)}. In Vivado HLS, all the inputs and outputs are in 32-bit fixed format and are assigned as AP\_NONE. The output Verilog Top file of VeriPy will be tested and simulated in Vivado 2024.1 to obtain performance and resource consumption data. 
        
        In addition to the five major SDR examples, several other simple designs have been developed and tested using different versions of Vivado HLS and various FPGA boards. All Python source files of VeriPy, generated Verilog top-level designs, comparison tables, and tool usage instructions have been uploaded to GitHub for public access. (https://github.com/RainChinChao/PythonToVerilogTool).

\begin{equation}
\label{eq1}
 MAC16(A,B) =\sum_{i=0}^{15} A(i) * B(i)
\end{equation}

\begin{equation}
\label{eq_cont}
 T =-\frac{2*pi*k*i}{n}
 \end{equation}

\begin{equation}
\label{eq2}
 FFT32\_real(k) =\sum_{i=0}^{31} Xreal(i) * cos (T) - Ximag(i) * sin (T)
 \end{equation}

\begin{equation}
\label{eq3}
 FFT32\_imag(k) =\sum_{i=0}^{31} Xreal(i) * sin (T) + Ximag(i) * cos (T)
\end{equation}

\begin{algorithm}[H]
\caption{BPSK\_Demodulation}\label{alg:Demodulation}
\begin{algorithmic}
\STATE 
\STATE {\textsc{BPSK\_Demodulation}}$(\mathbf{input\_sig[10]} )$
\STATE \hspace{0.5cm}$ $\textbf{ for } $ i = 1,...,10$
\STATE \hspace{1cm}$ t[i] \gets  i/Sample\_rate $
\STATE \hspace{1cm}$ carrier[i] \gets  cos(2 * pi * carrier\_freq * t[i]) $
\STATE \hspace{1cm}$ temp \gets  temp + carrier[i] * input\_sig[i] $

\STATE \hspace{0.5cm}$ $\textbf{ if } $    (\mathbf{temp} >= 0)$
\STATE \hspace{1cm}$ output \gets  1 $
\STATE \hspace{0.5cm}$ \textbf{ else } $
\STATE \hspace{1cm}$ output \gets  0 $

\end{algorithmic}
\label{alg1}
\end{algorithm}

\begin{algorithm}[H]
\caption{BPSK\_Modulation}\label{alg:Modulation}
\begin{algorithmic}
\STATE 
\STATE {\textsc{BPSK\_Modulation}}$(\mathbf{input\_bit} )$
\STATE \hspace{0.5cm}$ $\textbf{ for } $ i = 1,...,10$
\STATE \hspace{1cm}$ t[i] \gets  i/Sample\_rate $
\STATE \hspace{1cm}$ carrier[i] \gets  cos(2 * pi * carrier\_freq * t[i]) $
\STATE \hspace{1cm}$ carrier\_m[i] \gets  -cos(2 * pi * carrier\_freq * t[i]) $
\STATE \hspace{0.5cm}$ $\textbf{ if } $    (\mathbf{input\_bit} == 1)$
\STATE \hspace{1cm}$ bpsk\_s[i] \gets  carrier[i] $
\STATE \hspace{0.5cm}$ \textbf{ else } $
\STATE \hspace{1cm}$ bpsk\_s[i] \gets  carrier\_m[i] $

\end{algorithmic}
\label{alg2}
\end{algorithm}

\begin{algorithm}[H]
\caption{Back\_Propagation}\label{alg:BackPropagation}
\begin{algorithmic}
\STATE 
\STATE {\textsc{Back\_Propagation}}$(\mathbf{X[8], Y[4]} )$
\STATE \hspace{0.5cm}$ $\textbf{ for } $ i = 1,...,4$
\STATE \hspace{1cm}$ net\_input[i] \gets  \mathbf{X}[2*i]*weight[1] + \mathbf{X}[2*i+1] *weight[2]$
\STATE \hspace{1cm}$ output[i] \gets  1/2 + net\_input[i]/2 ($\textbf{ originally } $ 1/(1+exp(net\_input[i])) $
\STATE \hspace{1cm}$ error[i] \gets  \mathbf{Y}[i]-output[i] $
\STATE \hspace{1cm}$ sig\_deri\_out[i] \gets  (1-output[i])*output[i] $
\STATE \hspace{1cm}$ delta[i] \gets  error[i] * sig\_deri\_out[i] $

\end{algorithmic}
\label{alg3}
\end{algorithm}

    The five tested examples are illustrated in Equations \ref{eq1}, \ref{eq2}, \ref{eq3} (Equation \ref{eq_cont} is part of Equation \ref{eq2}), and Algorithms \ref{alg1}, \ref{alg2}, and \ref{alg3}. Equation \ref{eq1} represents a multiply-accumulate (MAC) operation, which is commonly extracted from SDR designs such as Automatic Modulation Classification (AMC), as well as CNN-based AMC implementations \cite{bib12, bib13, bib14}. Equations \ref{eq2} and \ref{eq3} describe components of a 32-point Fast Fourier Transform (FFT), a fundamental operation widely used in SDR applications. Algorithms \ref{alg1} and \ref{alg2} demonstrate the Binary Phase-Shift Keying (BPSK) modulation and demodulation algorithms, both of which have been designed and implemented using VeriPy and Vivado HLS. Algorithm \ref{alg3} presents the backpropagation function, which is commonly used in convolutional neural networks (CNNs), a technique increasingly applied in modern SDR systems.

\subsection{Perforamance and Resource Consumption}

Table \ref{tab:VeriPy2024Res}, Table \ref{tab:HLS2024Res}, and Table \ref{tab:HandVerilog2024Res} present the maximum operating frequency and resource utilisation of selected designs generated by VeriPy, Vivado HLS 2024, and hand-coded Verilog, respectively. Both the VeriPy-generated and hand-coded designs were synthesised and implemented using Vivado 2024.1 to ensure a consistent evaluation environment. P means pipelined design while U means unrolled design.

\begin{table}
    \centering
    \caption{Algorithm Performance and Resource Results From VeriPy run by Vivado 2024.1.}
    \begin{tabular}{|>{\centering\arraybackslash}p{2.3cm}|
                    >{\centering\arraybackslash}p{1.2cm}|
                    >{\centering\arraybackslash}p{0.6cm}|
                    >{\centering\arraybackslash}p{0.6cm}|
                    >{\centering\arraybackslash}p{0.6cm}|
                    >{\centering\arraybackslash}p{0.6cm}|
                   }
    \hline
        VeriPy & Max Freq (MHz) & LUT & FF & DSP & BRAM\\
    \hline
        MAC 16 P & 181 & 5079 & 2930 & 64 & 0\\
        \hline
        MAC 16 U & 895 & 4320 & 1024 & 64 & 0\\
        \hline
        FFT 32 imag P & 182 & 21388 & 9635 & 240 & 0\\
        \hline
        FFT 32 real P & 182 & 19539 & 9699 & 240 & 0\\
        \hline
        FFT 32 U & 180 & 32620 & 7796 & 464 & 0\\
        \hline
        Demodulation P& 119 & 3013 & 1453 & 40 & 0\\
        \hline
        Demodulation U & 181 & 2631 & 962 & 40 & 0\\
        \hline
        Modulation P & 211 & 108 & 51 & 0 & 0\\
        \hline
        Modulation U & inf & 9 & 1 & 0 & 0\\
        
        \hline
        Back Propagation P & 108 & 1389 & 516 & 16 & 0\\
        \hline
        Back Propagation U & 132 & 4639 & 1018 & 64 & 0\\
        \hline
    \end{tabular}
    
    \label{tab:VeriPy2024Res}
\end{table}		
		
	\begin{table}
    \centering
    \caption{Algorithm Performance and Resource Results From Vivado HLS 2024.1.}
    \begin{tabular}{|>{\centering\arraybackslash}p{2.3cm}|
                    >{\centering\arraybackslash}p{1.2cm}|
                    >{\centering\arraybackslash}p{0.6cm}|
                    >{\centering\arraybackslash}p{0.6cm}|
                    >{\centering\arraybackslash}p{0.6cm}|
                    >{\centering\arraybackslash}p{0.6cm}|
                   }
        \hline
        HLS & Max Freq (MHz) & LUT & FF & DSP & BRAM\\
        \hline
        MAC 16 & 776 & 108 & 77 & 4 & 0\\
        \hline
        MAC 16 U & 776 & 108 & 77 & 4 & 0\\
        \hline
        MAC 16 P & 362 & 198 & 234 & 8 & 0\\
        \hline
        FFT 32 & 259 & 122 & 142 & 8 & 0\\
        \hline
        FFT 32 U & 259 & 122 & 142 & 8 & 0\\
        \hline
        FFT 32 P & 109 & 9720 & 977 & 12 & 0\\
        \hline
        Demodulation & 195 & 705 & 279 & 15 & 0\\
        \hline
        Demodulation U& 205 & 184 & 139 & 4 & 0\\
        \hline
        Demodulation P & 504 & 201 & 166 & 0 & 0\\
        \hline
        Modulation & 178 & 677 & 323 & 7 & 0\\
        \hline
        Modulation U & 307 & 249 & 194 & 0 & 0\\
        \hline
        
        Modulation P & 1818 & 16 & 6 & 0 & 0\\
        \hline
        Back Propagation & 149 & 647 & 604 & 8 & 0\\
        \hline
        Back Propagation U & 149 & 647 & 604 & 8 & 0\\
        \hline
        Back Propagation P & 174 & 498 & 155 & 8 & 0\\
\hline
    \end{tabular}
    
    \label{tab:HLS2024Res}
\end{table}
		
\begin{table}
    \centering
    \caption{Algorithm Performance and Resource Results From Hand-coded Veriog run by Vivado 2024.1.}
    \begin{tabular}{|>{\centering\arraybackslash}p{2.3cm}|
                    >{\centering\arraybackslash}p{1.2cm}|
                    >{\centering\arraybackslash}p{0.6cm}|
                    >{\centering\arraybackslash}p{0.6cm}|
                    >{\centering\arraybackslash}p{0.6cm}|
                    >{\centering\arraybackslash}p{0.6cm}|
                   }
        \hline
        Verilog & Max Freq (MHz) & LUT & FF & DSP & BRAM\\
        \hline
        MAC 16 P & 119 & 4848 & 2413 & 64 & 0\\
        \hline
        MAC 16 U & 895 & 4320 & 1024 & 64 & 0\\
        \hline
        FFT 32 P & 181 & 15379 & 7758 & 192 & 0\\
        \hline
        FFT 32 U & 181 & 13024 & 3136 & 192 & 0\\
        \hline
        Demodulation P& 181 & 2969 & 1416 & 40 & 0\\
        \hline
        Demodulation U & 181 & 2607 & 609 & 40 & 0\\
        \hline
        Modulation U & 205 & 7 & 2 & 0 & 0\\
        \hline
        Modulation P & inf & 7 & 2 & 0 & 0\\
        \hline
        Back Propagation P & 200 & 127 & 319 & 8 & 0\\
      \hline  
        Back Propagation U & 471 & 51 & 1 & 4 & 0\\
\hline
    \end{tabular}
    
    \label{tab:HandVerilog2024Res}
\end{table}

As shown in Table \ref{tab:VeriPy2024Res}, Table \ref{tab:HLS2024Res}, and Table \ref{tab:HandVerilog2024Res}, VeriPy demonstrates superior performance in terms of maximum operating frequency compared to Vivado HLS 2024.1 for the unrolled designs of MAC16, FFT32, and modulation, achieving up to 1.7× higher frequency. When compared with hand-coded Verilog, VeriPy achieves comparable maximum frequencies. Regarding resource utilisation, VeriPy adopts a modular architecture with a top-level file structure, resulting in resource usage similar to that of hand-coded Verilog, though slightly higher than that of the Vivado HLS version.

\subsection{Readability and Optimisation}

		Figure \ref{GenCodeVeriPy} and Figure \ref{GenCodeHLS}  display the Verilog code generated separately by the Vivado HLS and VeriPy approach. In the Vivado HLS version, the code is scarcely readable, even for a hardware engineer. However, according to the Verilog code generated by VeriPy, with the aid of automatic comment generation highlighting each step, readability is greatly enhanced compared to HLS. This ensures the feasibility of further improving the performance or power of the SDR application at the hardware level, if necessary.
		
		Furthermore, the hardware design produced by this study consists of a package containing a generated Verilog Top file along with the provided Verilog files for fundamental functions. These fundamental function Verilog files can be easily substituted for performance or power optimisation purposes. VeriPy supports flexible optimisation at both the Python (high-level) and Verilog (low-level) stages. Users can either substitute the hardware implementation of fundamental functions or modify the overall hardware structure to optimise for performance, power, and area (PPA).
		
		As shown in Table~\ref{tab:OptiComp}, Equation~\ref{eq1} is selected to demonstrate high-level, low-level, and combined optimisations. High-level optimisation refers to reducing computation within the design—for example, by minimising the number of operations inside a loop. Low-level optimisation involves replacing a fundamental hardware function with a more efficient custom implementation. Combined optimisation applies both high-level and low-level techniques to achieve improved overall design efficiency.
		
\begin{table}
    \centering
    \caption{Performance and Resource Consumption Comparison between Optimised and Unoptimised MAC16.}
    \begin{tabular}{|>{\centering\arraybackslash}p{2.3cm}|
                    >{\centering\arraybackslash}p{1.2cm}|
                    >{\centering\arraybackslash}p{0.6cm}|
                    >{\centering\arraybackslash}p{0.6cm}|
                    >{\centering\arraybackslash}p{0.6cm}|
                    >{\centering\arraybackslash}p{0.6cm}|
                   }
        \hline
         & Max Freq & LUT & FF & DSP & BRAM\\
        \hline
        Unoptimised MAC16 & 181 & 5079 & 2930 & 64 & 0\\
        \hline
        Python Optimised MAC16 & 119 & 4325 & 995 & 64 & 0\\
        \hline
        Verilog Optimised MAC16 & 218 & 2789 & 2418 & 0 & 0\\
        \hline
        Python+Verilog Optimised MAC16 & 186 & 2245 & 995 & 0 & 0\\
        \hline
    \end{tabular}
    
    \label{tab:OptiComp}
\end{table}

    The results in Table~\ref{tab:OptiComp} demonstrate a frequency improvement of 1.2× and 1.03× when Verilog-level and combined Python+Verilog optimisations are applied, respectively. The table also highlights notable reductions in resource utilisation: Python-level optimisation yields a 15\% reduction in LUT usage and a 67\% reduction in flip-flop (FF) usage; Verilog-level optimisation results in a 45\% reduction in LUTs, 17\% in FFs, and complete elimination (100\% reduction) of DSP usage. When both Python and Verilog optimisations are applied, the design achieves a 56\% reduction in LUTs, 67\% in FFs, and 100\% in DSPs. These results confirm the effectiveness and flexibility of VeriPy in supporting both high-level and low-level design optimisations.

\subsection{Length of Code (LoC)}

	The length of the input code is considered one of the comparison aspects of design effort in a code generation tool, like in \cite{bib16}. Pragmas, which are commonly used in Vivado HLS for input/output port definitions and optimisation directives, represent another important factor when evaluating the ease of use of an HLS tool. Therefore, in this section, the lengths of the input code and the number of pragmas for the previous four examples are compared between VeriPy and HLS, as shown in Table {tab:lengthofCode}.
	
	The data in Table~\ref{tab:lengthofCode} shows that the VeriPy tool requires a comparable number of input code lines to Vivado HLS when the design is simple. This is primarily due to VeriPy’s one-to-one correspondence concept and its use of a limited set of fundamental functions, whereas Vivado HLS allows multiple operations to be expressed within a single statement and has more already-designed complicated functions that can be directly used. However, as VeriPy continues to evolve and its function library expands, the overall input code length is expected to decrease. Notably, VeriPy requires no pragma directives during design, further simplifying the development process.

\begin{table}
    \centering
    \caption{Length of Input Code Comparison Between VeriPy and Vivado HLS.}
    \begin{tabular}{|>{\centering\arraybackslash}p{2cm}|
                    >{\centering\arraybackslash}p{1cm}|
                    >{\centering\arraybackslash}p{1cm}|
                    >{\centering\arraybackslash}p{1cm}|
                    >{\centering\arraybackslash}p{1cm}|
                   }
        \hline
         & VeriPy &VeriPy Pragma & Vivado HLS & Vivado HLS pragma\\
         \hline
        MAC16 & 15 & 0 &13 &4\\
        \hline
        FFT32 & 67 & 0 & 24 & 7\\
        \hline
        Modulation & 41 & 0 & 32 & 7\\
        \hline
        Demodulation & 50 & 0 &34 & 6 \\
        \hline
        BackPropagation & 43 & 0&35&9\\
        \hline
    \end{tabular}
    
    \label{tab:lengthofCode}
\end{table}

\subsection{Performance and Resource Estimation}	
	
VeriPy also supports performance estimation of the number of processing cycles in pipelined designs, as well as resource estimation for both unrolled and pipelined designs.

As shown in Table~\ref{tab:VeriPyResEsti}, the performance estimation for processing cycles is 100\% accurate. The resource estimation, while not exact, closely approximates the actual resource utilisation observed during the implementation stage.

    \begin{table}
    \centering
    \caption{Estimated Resource by VeriPy.}
    \begin{tabular}{|>{\centering\arraybackslash}p{2cm}|
                    >{\centering\arraybackslash}p{1cm}|
                    >{\centering\arraybackslash}p{1cm}|
                    >{\centering\arraybackslash}p{0.8cm}|
                    >{\centering\arraybackslash}p{0.6cm}|
                    >{\centering\arraybackslash}p{0.8cm}|}
        \hline
        VeriPy & Process Cycle & LUT & FF & DSP & BRAM\\
        \hline
        MAC 16 P & 17 & 5744 & 2930 & 64 & 0\\
        \hline
        MAC 16 U & NA & 4320 & 1024 & 64 & 0\\
        \hline
        FFT 32 P & 34 & 47968 & 8192 & 512 & 0\\
        \hline
        FFT 32 U & NA & 47968 & 8192 & 512 & 0\\
        \hline
        Demodulation P& 13 & 3752 & 673 & 40 & 0\\
        \hline
        Demodulation U & NA & 3800 & 705 & 40 & 0\\
        \hline
        Modulation P & 2 & 1140 & 10 & 0 & 0\\
        \hline
        Modulation U & inf & 820 & 10 & 0 & 0\\
        \hline
        Back Propagation P & 7 & 1921 & 320 & 20 & 0\\
        \hline
        Back Propagation U & NA & 7492 & 1152 & 80 & 0\\
        \hline
    \end{tabular}
    
    \label{tab:VeriPyResEsti}
\end{table}

\section{Conclustion}

		In this paper, VeriPy, a Python-based tool for SDR hardware accelerator development, has been proposed. The tool enables automatic generation of unrolled and pipelined hardware designs in Verilog based on high-level Python SDR algorithms, requiring no prior hardware design experience. VeriPy also supports automatic testbench generation, an extensible hardware library, performance and resource estimation, and enhanced readability of the generated Verilog code. Additionally, the tool offers strong potential for both high-level and low-level design optimisation. These features not only accelerate the design process but also lower the learning curve and bridge the gap between SDR software developers and hardware engineers.
        
        Performance and resource evaluation of VeriPy-generated designs—tested using identical SDR algorithms—shows that VeriPy can achieve up to 1.7× higher operating frequency than designs generated by Vivado HLS with optimisation pragmas, with only a minimal increase in resource usage. This improvement can be attributed to VeriPy’s one-to-one module mapping approach, which contrasts with the more abstract generation strategy used in Vivado HLS. Moreover, VeriPy’s strong optimisation potential allows for further PPA improvements when enhanced fundamental function modules or optimised architectural structures are integrated. Compared to hand-coded Verilog designs, VeriPy demonstrates comparable performance and resource efficiency, highlighting the quality of its hardware generation.

	One limitation to be addressed in future work is the need for users to follow a VeriPy-compatible coding style in the Python input. Although no hardware knowledge is required, the current syntax and structure rules may still slow down the design process. Future development should focus on relaxing these input constraints to further streamline design and enhance usability.

	In conclusion, the proposed Python-based VeriPy tool offers a compelling solution for SDR hardware accelerator generation. It excels in design speed, code readability, optimisation flexibility, and ease of use—thanks to features such as automatic testbench generation and accurate performance/resource estimation. While it may exhibit slightly higher resource usage than Vivado HLS, it delivers hardware quality comparable to hand-coded Verilog and remains accessible to users without hardware expertise, making it a practical and efficient tool for modern SDR hardware development.

\section{References Section}

\newpage

	\begin{IEEEbiography}[{\includegraphics[width=1in,height=1.25in,clip,keepaspectratio]{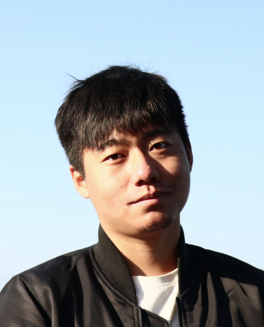}}]
		{YUQIN ZHAO,}
		received the B.E. degree in electrical and computer engineering from Tamkang University, New Taipei City, Taiwan, in 2021 and the MSc. degree in electronics from Nanyang Technological University, Singapore, in 2022. He is currently pursuing a Ph.D. degree in electronic and electrical engineering at the University of Sheffield, Sheffield, UK. 
		His research interest includes the development of an FPGA accelerator for AI and SDR, SDR implementation of SDR application, and high-level synthesis environment for SDR and AI applications on FPGAs.
		
	\end{IEEEbiography}

	\begin{IEEEbiography}[{\includegraphics[width=1in,height=1.25in,clip,keepaspectratio]{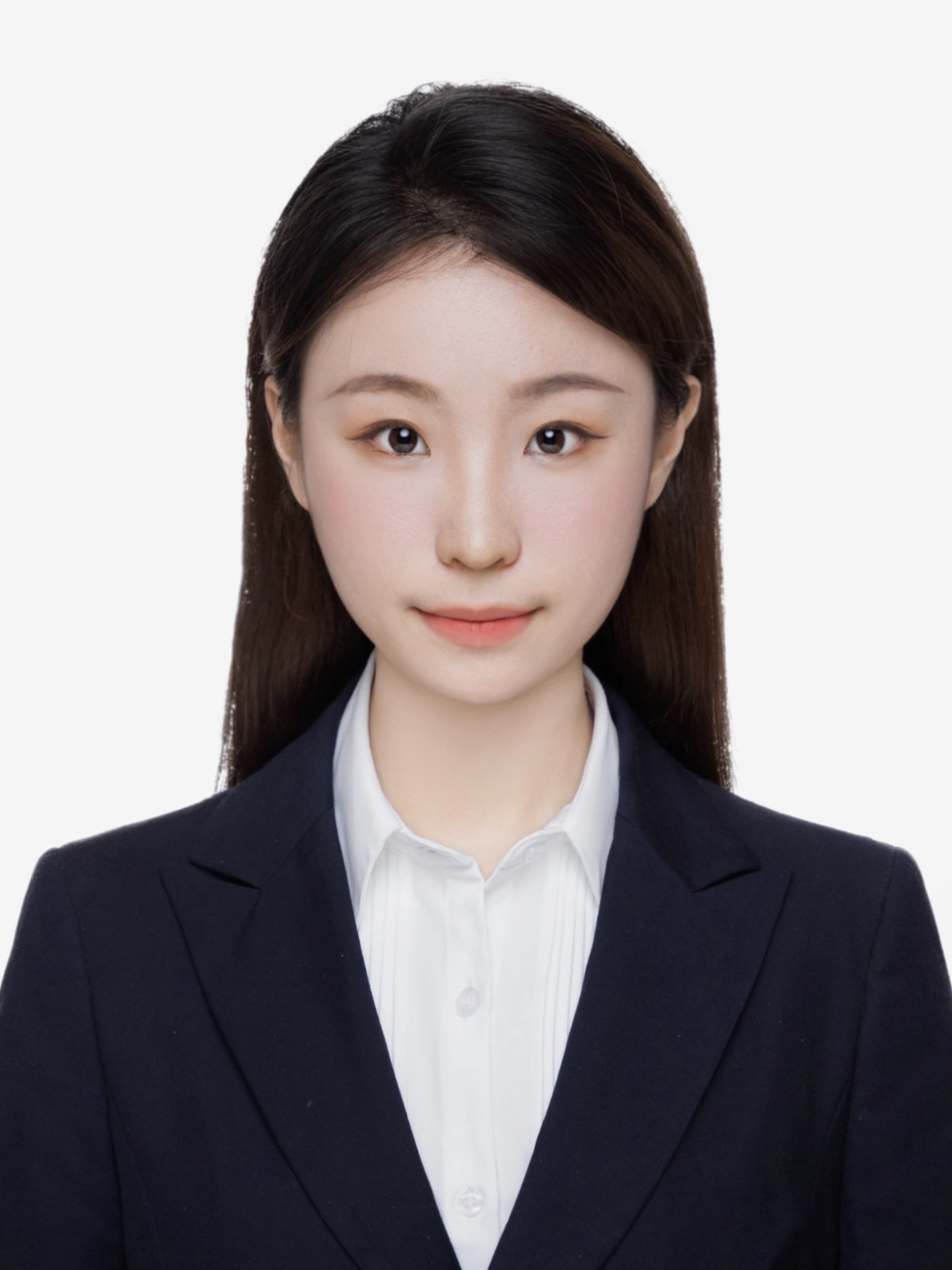}}]
		{YE LINGHUI} received the B.Eng. degree in Electronics and Computer Engineering from the University of Sheffield, UK, in 2025. She has been admitted to the Master of Computer Science  at the University of Sydney, Australia.

	\end{IEEEbiography}


\begin{IEEEbiography}[{\includegraphics[width=1in,height=1.25in,clip,keepaspectratio]{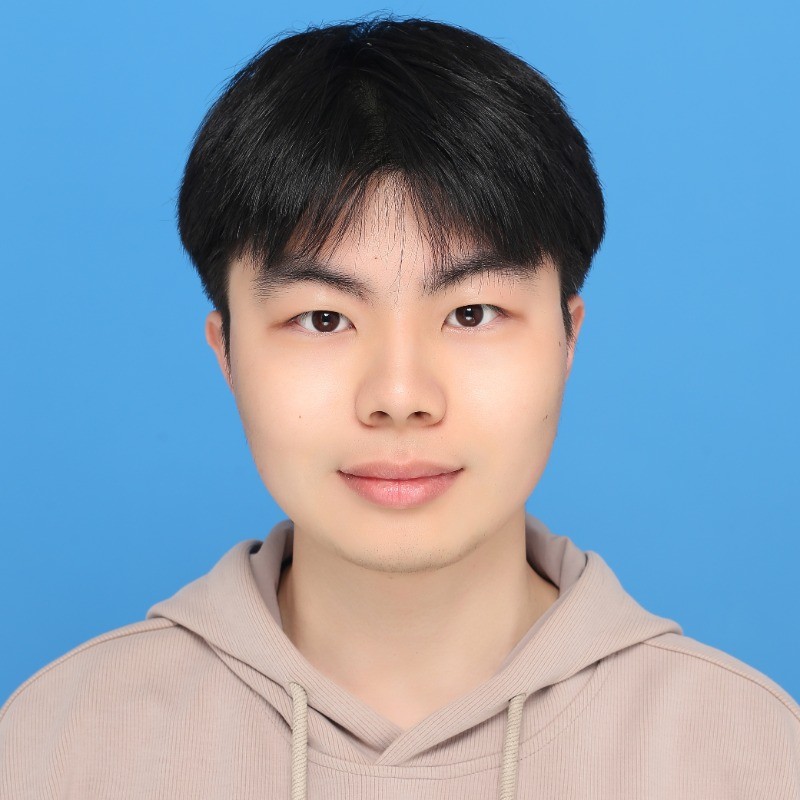}}]{Haihang Xia}
received the MSc degree in Electronic and Electrical Engineering from University of Glasgow, Glasgow, United Kingdom, in 2023. He is currently pursuing the PhD degree in Electronic and Electrical Engineering with the
School of Electronic and Electrical Engineering, University of Sheffield, Sheffield, United Kingdom. His research interests include Spiking Neural Networks, Neuromorphic computing, AI acceleration, and digital systems design.
\end{IEEEbiography}

	\begin{IEEEbiography}[{\includegraphics[width=1in,height=1.25in,clip,keepaspectratio]{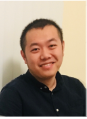}}]
		{Dr Tiantai Deng,}
		received his PhD from Queen’ s University Belfast in 2019, and BEng from Harbin Institute of Technologies in 2015. He is currently a lecturer at the University of Sheffield. 
		Prior to his career as an academic, he was a senior engineer at HiSilicon, Huawei. His main research focus is on hardware acceleration for image processing, deep learning, and high-level design environments.
	\end{IEEEbiography}

	\begin{IEEEbiography}[{\includegraphics[width=1in,height=1.25in,clip,keepaspectratio]{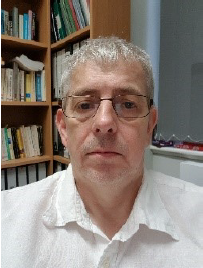}}]
		{Luke Seed,}
		is a Senior Lecturer in the Department of Electronic and Electrical Engineering at the University of Sheffield. His research interests extend from chip design through to holographic lithography.
	\end{IEEEbiography}

\vfill

\end{document}